\documentclass[prb,aps,twocolumn,showpacs,floatfix]{revtex4-2}

\usepackage{amsfonts}
\usepackage{amsmath}
\usepackage{bbm}
\usepackage{hyperref}
\usepackage{graphicx}
\usepackage{dcolumn}
\usepackage{bm}
\usepackage{subfigure}
\usepackage{amssymb}
\hypersetup{colorlinks,linkcolor=blue,citecolor=blue,filecolor=blue,urlcolor=blue}
\newcommand{\me}{\mathrm{e}} \newcommand{\mi}{\mathrm{i}}\newcommand{\piup}{\mathrm{\pi}}
\newcommand{\dif}{\mathop{}\!\mathrm{d}}
\renewcommand{\vec}[1]{\bm{#1}}
\DeclareMathAlphabet{\mathfsl}{OT1}{cmss}{m}{sl}

\begin{document}

\title{Topological Quantum Phase Transitions in Metallic Shiba Lattices}
\author{Ning Dai$^{1,2}$}
\author{Kai Li$^2$}
\author{Yan-Bin Yang$^{2,3}$}
\author{Yong Xu$^{1,2}$}
\email{yongxuphy@tsinghua.edu.cn}
\affiliation{$^{1}$Shanghai Qi Zhi Institute, Shanghai 200030, People's Republic of China}
\affiliation{$^{2}$Center for Quantum Information, IIIS, Tsinghua University, Beijing 100084, People's Republic of China}
\affiliation{$^{3}$Department of Physics, Hong Kong University of Science and Technology, Clear Water Bay, Hong Kong 999077, People's Republic of China}
\begin{abstract}
Shiba bands formed by overlapping Yu-Shiba-Rusinov subgap states in magnetic impurities on a superconductor
play an important role for realizing topological superconductors.
Here, we theoretically demonstrate the existence of topological gapless Shiba bands on a magnetically doped $s$-wave superconducting surface with Rashba spin-orbit coupling in the presence of a weak in-plane magnetic field.
Such bands develop from gapped Shiba bands through Lifshitz phase transitions
accompanied by second-order quantum phase transitions for the intrinsic thermal Hall conductance.
We also find a mechanism in Shiba lattices that protects the first-order quantum phase transitions
for the intrinsic thermal Hall conductance.
Due to the long-range hopping in Shiba lattices, the topological Shiba metal exhibits intrinsic thermal Hall
conductance with large nonquantized values. As a consequence,
there emerge a large number of second-order quantum phase transitions.
\end{abstract}
\maketitle
\section{introduction}
Topological superconductors have attracted a great amount of attention during the last decade due to their
potential applications in topological quantum computation~\cite{Green PRB2000,Kitaev Phys.Usp.2001,Ludwig PRB2008,Kitaev 2009,DasSarma PRL2010,Alicea Rep.Prog.Phys.2012,Beenakker 2013,Ando Rep.Prog.Phys.2017,Wendin Rep.Prog.Phys.2017,Ferrini PRL2020}.
In the context, Shiba lattices
play an important role, since they provide a versatile tool
to realize highly controllable topological superconducting phases~\cite{Oppen PRB2013,Ojanen PRB2014,Sau PRB2015,Ojanen PRL2015,Ojanen PRB2016,schecter2016self,li2016two,Ojanen nat.commun2018}.
Shiba lattices are formed by overlapping Yu-Shiba-Rusinov (YSR) subgap states, bound states
occurring in magnetic impurities when placed on a superconducting surface~\cite{Shiba 1968,Zhu Rev.Mod.Phys.2006,yao2014enhanced,refereeAadded,hatter2015magnetic,ruby2016orbital,yang2020observation,ding2021tuning,beck2021spin,wang2021PRL}.
Such lattices can be utilized to realize topological superconductivity with high Chern numbers
due to the long-range hopping between two YSR subgap states~\cite{Ojanen PRL2015}.
Remarkably,
a very recent
experiment reports on an observation of topological Shiba bands in a magnet-superconductor hybrid system~\cite{Wiebe2021NP}.
Topological phases can not only exist in gapped systems but also in gapless systems~\cite{VishwanathRMP,XuReview}.
However, for Shiba lattices,
previous studies either focused on
gapped superconductors in regular Shiba lattices~\cite{Ojanen PRL2015,Ojanen PRB2016} or gapless superconductors (but Anderson localized)
in Shiba glasses~\cite{Ojanen nat.commun2018}.
Although
it has been theoretically predicted that topological metals can emerge in fermionic superfluids in
cold atom systems~\cite{Xu PRL2014,Xu PRL2015,HuPRL2014,Xu2015PRLBKT,Kamenev PRL2018},
it is unclear whether topological metals can arise from the subgap band formed by
the YSR states.

\begin{figure}
	\includegraphics[width=\linewidth]{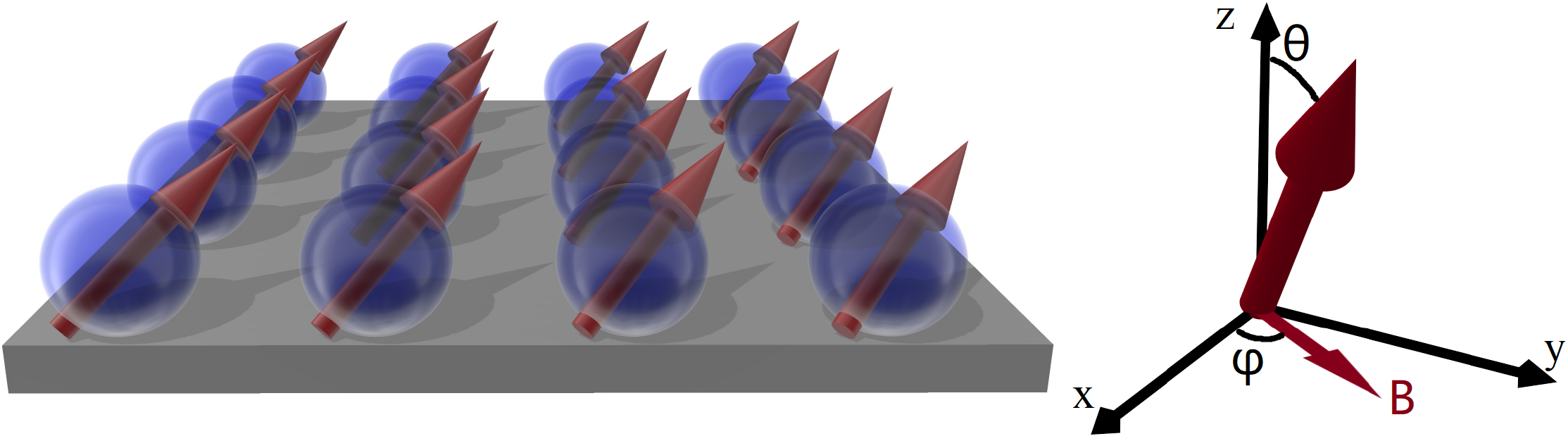}
	\caption{(Color online)
		Ferromagnetic impurities are arranged into a square lattice on an $s$-wave superconducting surface with Rashba spin-orbit coupling,
		resulting in topological metallic phases in Shiba bands.
		The direction of the magnetization (specified by the polar angle $\theta$ and azimuthal angle $\varphi$) deviates from
		the surface normal direction ($z$ direction) due to the presence of an in-plane magnetic field $\vec{B}$.
	}
	\label{fig1}
\end{figure}

Motivated by the experimental progress in topological Shiba bands, we here study the topological phases in a two-dimensional (2D) lattice formed by ferromagnetic impurities
on a 2D $s$-wave superconducting surface with Rashba spin-orbit coupling.
We theoretically predict the emergence of topological metallic phases in Shiba bands
in the presence of a weak in-plane magnetic field, which drives the direction of the magnetic moment away from a surface normal vector.
Starting from a gapped topological superconducting phase, one can obtain the metallic phase through a
Lifshitz phase transition by varying a system parameter, such as the Fermi wavevector or the spin-orbit coupling strength.
The transition also manifests in a second-order quantum phase transition for the intrinsic thermal Hall conductance.
In addition,
it has been shown that owing to particle-hole symmetry, the intrinsic thermal Hall conductance always exhibits a
first-order quantum phase transition if an energy gap closes at a high-symmetry momentum~\cite{Kamenev PRL2018}.
When the energy gap closing points deviate from high-symmetry momenta, the first-order quantum phase transition is
not protected in a metallic phase since these points are usually not pinned at zero energy.
Remarkably, we find abundant first-order quantum phase transitions in Shiba metals arising from the
energy gap closing at \textit{non-high-symmetry} momenta.
We demonstrate a new mechanism (called reciprocal lattice reflection symmetry) in Shiba lattices that
fixes the band touching point at zero energy and thus protects the first-order phase transition.
Moreover, we illustrate that the topological metals exhibit intrinsic thermal Hall conductance
with large nonquantized values due to the long-range hopping supported by the YSR states,
leading to many continuous quantum phase transitions.

\section{Model}
The proposed Shiba metal is hosted on a magnetic impurity lattice deposited on a superconducting surface (see Fig.~\ref{fig1}).
Each impurity binds a YSR subgap state, which couples with other YSR states and forms a Shiba lattice.
The coupling between two YSR subgap states depends on the direction of the corresponding magnetic impurities.
Previous studies focus on the case where impurities constitute a ferromagnetic phase with the direction of the magnetization being perpendicular to the superconducting surface.
However,
such a Shiba lattice respects both a two-fold rotational symmetry and particle-hole symmetry, which rule out the metallic phase~\cite{Ojanen PRL2015,Ojanen PRB2016}.
For this reason, topological Shiba metals can only be found in Shiba lattices with tilted magnetization, where the magnetic moments of impurities deviate from the normal vector of the surface.

\begin{figure}
\includegraphics[width=\linewidth]{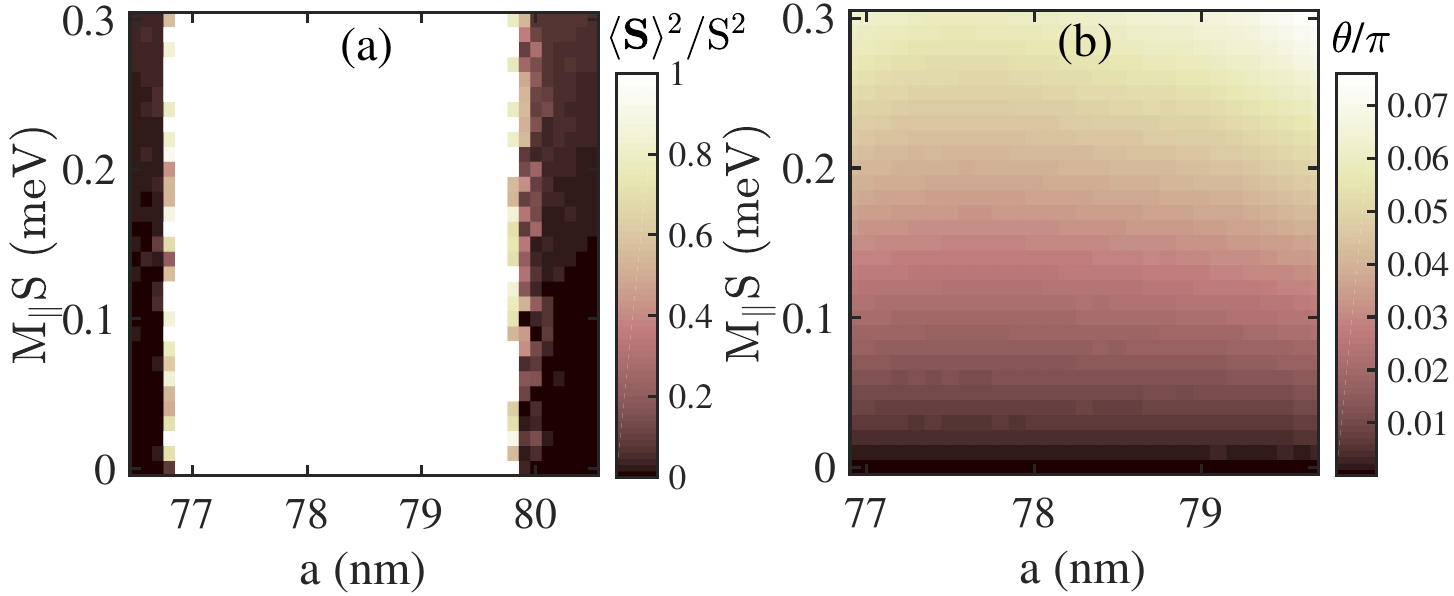}
\caption{(Color online) Magnetic phase diagram with respect to the Shiba lattice constant $a$ and the Zeeman field $M_\parallel S$
	mapped out based on (a) the renormalized magnetization amplitude $\langle {\bf S}\rangle^2/S^2$ and
(b) the magnetization direction characterized by the polar angle $\theta/\pi$.
The ferromagnetic phase is identified as $\langle {\bf S}\rangle^2/S^2=1$, where the magnetic polar angle is able to reach $0.06\piup$ within $M_\parallel S<0.3$ meV.
Here, we take $D S^2/2=1$~meV, the Fermi wavevector $k_F=1.57\times10^8~\mathrm{m}^{-1}$, the Fermi energy $E_F=94$~meV,
the renormalized Rashba coefficient $\alpha_R/v_F=0.07$ ($v_F$ is the Fermi velocity) and the azimuthal angle $\varphi=0$
for a $20\times20$ impurity lattice.
}
\label{fig2}
\end{figure}

To see whether the impurities are able to support tilted magnetization, we consider the following classical Hamiltonian of the magnetic impurities
under a weak in-plane magnetic field~\cite{Utsumi2004PRB},
\begin{equation}
\label{HS}
H_\mathrm{S}=\sum\nolimits_i(-\frac{D}{2}S_{i,z}^2 + M_\parallel S_{i,\parallel}) + H_\mathrm{RKKY}.
\end{equation}
Here $D$ represents the strength of the crystal field anisotropy, $M_\parallel$ denotes the coupling strength between the classical spin $\vec{S}_i$ and an external in-plane
magnetic field $\vec{B}=B \vec{e}_{\varphi\parallel}$, $S_{i,\parallel} = \vec{S}_i \cdot \vec{e}_{\varphi\parallel}$ denotes the in-plane component of the spin, and $H_\mathrm{RKKY}$ is
the Ruderman-Kittel-Kasuya-Yosida (RKKY) Hamiltonian of the impurities under Rashba spin-orbit coupling~\cite{Utsumi2004PRB} (see Appendix A for details).
In the absence of RKKY interactions, the magnetic moment of each impurity atom prefers the surface normal $z$-axis when $D>0$;
an in-plane magnetic field would change its direction.
When the RKKY interaction is included, we use the Monte Carlo method to investigate the ground state properties of $H_\text{S}$ in Eq.~(\ref{HS}).
We remarkably find the existence of a ferromagnetic phase with tilted magnetization, as shown in Fig.~\ref{fig2}(a) by a white regime.
The tilted angle $\theta$ in the ferromagnetic regime can be enlarged by increasing the Zeeman field $M_\parallel S$ [see Fig.~\ref{fig2}(b)].
All these parameters are in reasonable scales, implying an experimentally accessible ferromagnetic Shiba lattice with tilted magnetization.
{\color{red}
We note that the impurities can be arranged on the superconducting surface by a scanning tunneling microscope (STM)~\cite{IBM1990NATURE}.}

With a ferromagnetic phase for impurities, we now turn to study the properties of the YSR states.
Since any spin flip of an individual impurity atom is suppressed by its ferromagnetic neighbors via RKKY interactions,
we assume that the classical spin model is valid in such a ferromagnetic regime~\cite{Ojanen PRL2015,heimes2015interplay}.
In this condition, the electronic Hamiltonian can be written as
\begin{equation}
\label{He}
\begin{aligned}
H_e =\zeta\tau_z &+ \alpha_R(\vec{\sigma}\times\vec{k})_z\tau_z+\Delta\tau_x+m_\parallel\sigma_\parallel
\\&  -J\sum\nolimits_i (\vec{S}_i \cdot \vec{\sigma}) \delta(\vec{r}-\vec{r}_i),
\end{aligned}
\end{equation}
where $\zeta={k^2}/({2m})-E_F$ denotes the kinetic energy of free electrons measured relative to the Fermi energy $E_F$
with $m$ being the effective mass, and $\Delta$ denotes the superconducting order parameter.
The Pauli matrices $\sigma$ and $\tau$ are defined on the spin and particle-hole subspaces, respectively.
An impurity $i$ is treated as a classical spin $\vec{S}_i$ localized at $\vec{r}_i$, coupled to the bulk electrons with the exchange coupling strength $J$.
Such a spin binds a YSR subgap state with eigenenergy $\Delta ({1-\alpha^2})/({1+\alpha^2})$ determined by a dimensionless coefficient $\alpha=mJS/2$.
A deep-in-gap YSR state occurs when $\alpha\approx1$.
In the presence of multiple impurities, the corresponding YSR states couple with each other and constitute a Shiba lattice.
The second term in Eq.~(\ref{He}) describes the Rashba spin-orbit coupling, an essential term to create nontrivial topology in Shiba lattices.
The in-plane magnetic field leads to a Zeeman splitting term $m_\parallel\sigma_\parallel$ for electrons, which seems to disturb the YSR state.
To suppress such an effect, the magnetic field is limited to a relatively weak scale.
For instance,
consider the Zeeman splitting $M_\parallel S$ of a magnetic atom below $0.3$~meV [see Fig.~\ref{fig2}].
If a magnetic moment of an impurity atom is five times larger than that of a free electron, then the
Zeeman splitting $m_\parallel$ can be limited to $0\sim0.06$~meV,
which is much smaller than the superconducting gap $\Delta\sim1$~meV.
In Appendix B, we show that such a weak Zeeman term is negligible.

For an isolated impurity, the YSR state is described
by $|+\uparrow\rangle$ and $|-\downarrow\rangle$ in the Nambu representation~\cite{Oppen PRB2013,refereeAadded}, where $|\tau\rangle$ ($\tau=\pm$)
and $|\sigma\rangle$ ($\sigma=\uparrow,\downarrow$)
are the eigenstates of $\tau_x$ and $\sigma_i = (\vec{S}_i /S) \cdot \vec{\sigma}$, respectively.
We derive a $2\times2$ tight-binding Hamiltonian to describe the low energy behavior of Shiba lattices
by projecting $H_e$ on these YSR states (see Appendix B for details),
\begin{equation}
H(\vec{r})=d_0(\vec{r})+\bm{d}(\vec{r})\cdot\bm{\sigma}.\label{Hamr}
\end{equation}
Here $H(\vec{r})$ represents the hopping matrix between two impurities with a displacement vector $\vec{r}=(r,\psi_{\bm r})$ in polar coordinates, and
for $\vec{r}\neq 0$,
\begin{eqnarray}
d_0(\vec{r})&=&\mi{\Delta}\mathrm{Im}A(r)\sin\theta\sin(\varphi-\psi_{\bm r})/2\\
d_x(\vec{r})&=&\mi{\Delta}\mathrm{Re}A(r)\cos\theta\sin(\varphi-\psi_{\bm r})/2\\
d_y(\vec{r})&=&\mi{\Delta}\mathrm{Re}A(r)\cos(\psi_{\bm r}-\varphi)/2\\
d_z(\vec{r})&=&-{\Delta}\mathrm{Re}S(r)/2,
\end{eqnarray}
where $S(r)$ and $A(r)$ are special functions composed of Bessel functions, which decay as $e^{-r/\xi}/r^{1/2}$, indicating that each YSR state couples to a number of other YSR states when $a<\xi$. Here $\xi=v_F/\Delta$ is the superconducting coherence length, and $v_F=k_F/m$ is the Fermi velocity. According to the values of other relevant parameters such as $k_F$, $E_F$ and $\Delta$,
we here set $\xi=1200$ nm. This long-range hopping nature lays the foundation for the unique topological property in Shiba lattices.
At $\vec{r}=0$, $d_z(0)=\Delta(1-\alpha^2)/(1+\alpha^2)$ and $d_{0,x,y}(0)=0$.
In the case with the magnetization aligning along $z$ (i.e., $\theta=0$),
Eq.~(\ref{Hamr}) reduces to the traditional case, which has been widely explored~\cite{Ojanen PRL2015,Ojanen PRB2016,Ojanen nat.commun2018}. This effective tight-binding Hamiltonian is valid in the low energy regime $E/\Delta\ll1$. When the YSR state is deep in the gap as $\alpha\approx1$, the majority of the Shiba band is in the low energy regime. For this reason, the effective Hamiltonian is appropriate to investigate the Shiba metal.
In momentum space, the Hamiltonian reads
\begin{equation}
H(\vec{k})=d_0(\vec{k})+\bm{d}(\vec{k})\cdot\bm{\sigma}\label{Hamk},
\end{equation}
where $d_{0,x,y}(\vec{k})$ and $d_z(\vec{k})$ are odd and even functions with respect to $\bm k$, respectively,
due to the particle-hole symmetry, i.e.,
$P^{-1}H(\vec{k})P=-H(-\vec{k})$ with $P=\sigma_x \kappa$
and $\kappa$ being the complex conjugate operator.
One can clearly see that a metallic phase cannot appear due to the vanishing of $d_0$ when $\theta=0$.

\begin{figure}
	\includegraphics[width=2.4in]{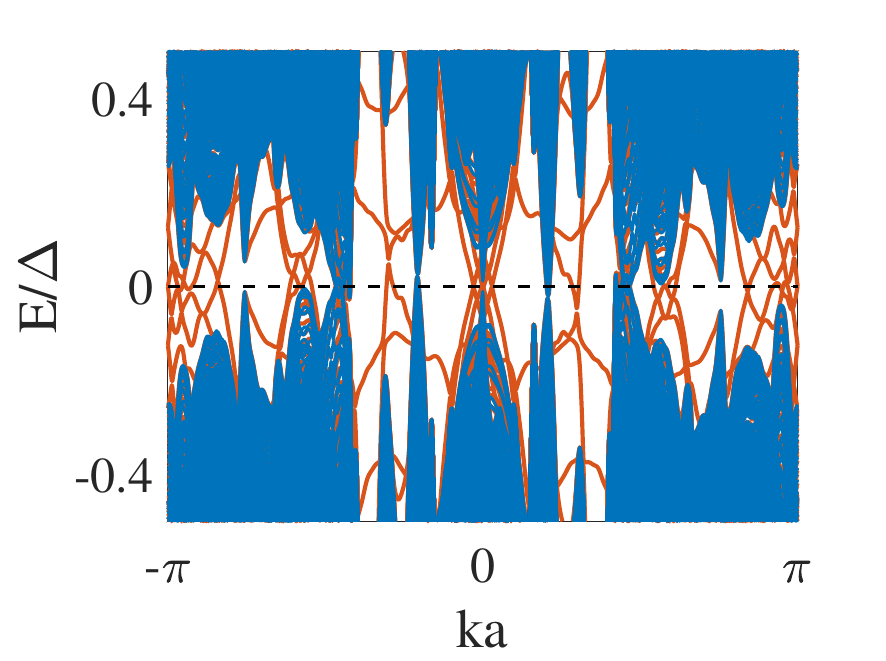}
	\caption{The energy spectrum of a 1000-atom wide Shiba lattice ribbon. The blue bands represent the bulk states, and the red lines describe the topologically protected edge states. Since the bulk states cross the Fermi surface (the black dashed line), this Shiba lattice is in a metallic phase. Here $k_F=1.524\times10^8\,m^{-1}$, $a=80$~nm and $\theta=0.04\pi$ corresponding to $M_\parallel S\sim0.2$~meV and $\varphi=0.25\piup$. Other parameters are the same as those in Fig.~\ref{fig2}. The ferromagnetism has been verified with these parameters.}
	\label{figR1}
\end{figure}

One may ask whether the quasiparticle excitation spectrum can exhibit a metallic phase.
The answer is affirmative. For simplicity, we first consider the nearest-neighbor hopping terms, which dominate,
and neglect other long-range hopping ones.
In this case, considering that an energy gap closes at $\vec{k}=0$, one can easily find that the eigenenergies of $H({\bm k})$
near the band touching point
can be approximated by $E({\bm k})\approx ka\Delta [\text{Im}A(a)\sin(\theta)\sin(\psi_{\bm k}-\varphi)\pm |\text{Re}A(a)|\sqrt{1-\sin^2(\theta)\sin^2(\psi_{\bm k}-\varphi)}]$
if $k_x a\ll 1$ and $k_y a\ll 1$ with ${\bm k}=(k,\psi_{\bm k})$ in polar coordinates.
If the absolute value of the first term is larger than that of the second term,
then the energies of both bands can become negative at some $\vec{k}$~\cite{Xu PRL2015}.
For example, when $\theta=\pi/2$, it requires that
$\tan|\psi_{\bm k}-\varphi|>|\text{Re}A(a)/\text{Im}A(a)|$, which can always be satisfied if $\text{Im}A(a)\neq 0$.
In a realistic case, a very small $\theta$ is able to render the energy spectrum gapless (see Fig.~\ref{figR1}).

\begin{figure}
\includegraphics[width=\linewidth]{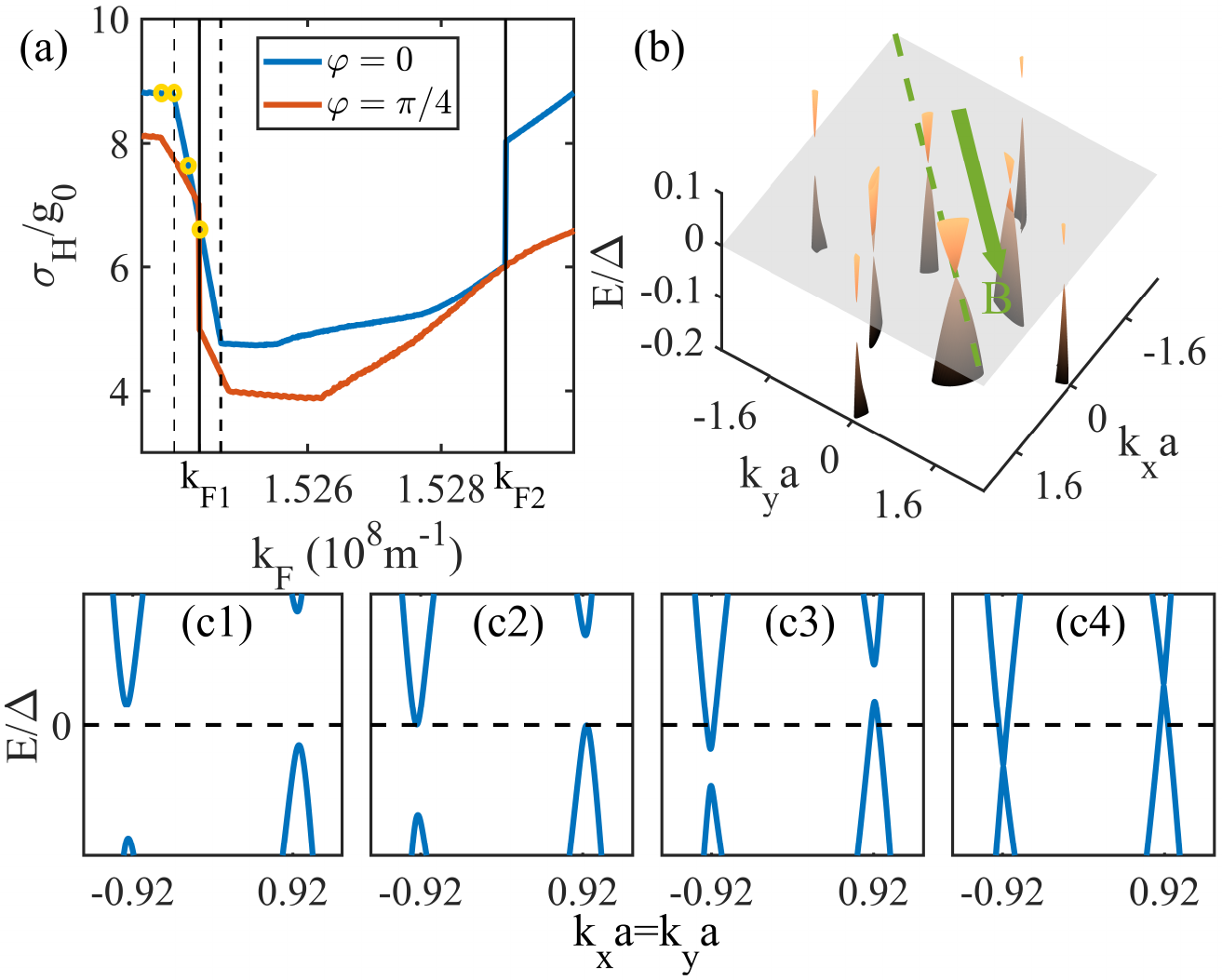}
\caption{(Color online) (a)~Intrinsic thermal Hall conductance $\sigma_H$ with respect to the Fermi wavevector $k_F$.
The reciprocal lattice reflection symmetry protected first-order topological quantum phase transitions happen at $k_F=k_{F1}$ when $\varphi=\piup/4$ or $k_F=k_{F2}$ when $\varphi=0$.
The first-order topological phase transition at $k_{F1}$ develops into two second-order topological phase transitions (highlighted by dashed lines)
when $\varphi$ changes from $\piup/4$ to 0.
(b)~Energy spectra around zero energy at $k_F=k_{F1}$ when $\varphi=\piup/4$.
The reciprocal lattice reflection symmetry guarantees that the valence and conduction bands on a symmetry line
${\bm k}=k{\bm e}_{\varphi\parallel}$ (green dashed line)	
can only close its energy gap at zero energy,
as visualized by two touching points on this line.
(c1)-(c4) Evolution of energy spectra along $k_x=k_y$ as $k_F$ changes from the left yellow circle to the right one in (a), illustrating
the development of a metallic phase through a Lifshitz phase transition.
Here, we set $a=80$~nm, $\theta=0.04\pi$ corresponding to $M_\parallel S\sim0.2$~meV, and
other parameters are the same as those in Fig.~\ref{fig2} so that
the ferromagnetism is guaranteed throughout the variation of $k_F$.
}
\label{fig3}
\end{figure}

The topological features of the metallic phase can be characterized by
the intrinsic thermal Hall conductance,
\begin{equation}\label{Hall}
\sigma_H=\frac{g_0}{2\piup}\sum\nolimits_n\int_\mathrm{BZ}{\dif^2\vec{k}}f[E_{n}(\vec{k})]\Omega_{n}(\vec{k}),
\end{equation}
where $\Omega_{n}$ denotes the Berry curvature of the $n$th band ($n=1,2$ refer to the valence and conduction bands, respectively), BZ stands for the first Brillouin zone,
$f(E)$ is the Fermi-Dirac distribution function, and $g_0={\piup^2k_\mathrm{B}^2T}/({6h})$ is the thermal conductance quantum with $T$ being the temperature
and $k_B$ being the Boltzmann constant~\cite{Greiner PRL1997}.
For a gapped system with temperatures much lower than the band gap, this thermal Hall conductance is equal
to the Chern number multiplied by $g_0$ due to the fully occupied valence band.
However, in the metallic regime, the intrinsic thermal Hall conductance is no longer quantized to an integer multiple of $g_0$ since both bands are partially occupied [see Fig.~\ref{fig3}(c3)].

If we only focus on the lowest band, which is separated from the higher band in momentum space,
then such band can has quantized nonzero Chern number, which leads to edge states (see the red lines in Fig.~\ref{figR1}),
illustrating the topological properties of the metallic phase.

\section{topological quantum phase transitions}
A topological phase transition occurs when the energy gap between the valence and conduction bands closes.
In the traditional gapped case without external magnetic fields, $d_0$ vanishes so that the energy gap can only close at zero energy, leading to
a quantized jump in $\sigma_H$ across the phase transition point. However, with nonzero $d_0$, the energy gap can close at nonzero energy,
in which case such a first-order quantum phase transition does not happen. Fortunately, the particle-hole symmetry guarantees that $d_0$ is an odd function with
respect to $\bm k$ so that $d_0$ has to vanish at high-symmetry momenta such as $(k_x a,k_y a)=(0,0),(0,\pi),(\pi,0),(\pi,\pi)$.
As a result, the first-order phase transition will take place if there is an energy gap closing at these high-symmetry points~\cite{Kamenev PRL2018}.

In Fig.~\ref{fig3}(a), we indeed observes the appearance of sharp changes in the thermal Hall conductance $\sigma_H$
as we vary $k_F$, revealing the first-order topological quantum phase transitions.
For example, when $\varphi=\pi/4$, $\sigma_H$ experiences a quantized decline at $k_{F1}$, indicating a phase transition between two topologically distinct metallic phases. However,
the energy spectra at $k_{F1}$ do not exhibit a gap closing at high-symmetry momenta [see Fig.~\ref{fig3}(b)].
Instead, two gap closings occur at momenta along $k_x=k_y$ at zero energy.
We show that the gap closings are protected by a reflection symmetry of the reciprocal lattices about the direction of the magnetic field.
Such a symmetry
ensures that $\mathcal{M}\mathcal{K}=\mathcal{K}$ where $\mathcal{K}$ is a set consisting of all reciprocal lattice vectors.
$\mathcal{M}$ is a reflection operator that acts on a reciprocal lattice vector
$\bm K=K_{\varphi\parallel}{\bm e}_{\varphi\parallel}+K_{\varphi\perp}{\bm e}_{\varphi\perp}$ resulting
in $\mathcal{M}{\bm K}=K_{\varphi\parallel}{\bm e}_{\varphi\parallel}-K_{\varphi\perp}{\bm e}_{\varphi\perp}$ with
${\bm e}_{\varphi\parallel}={\bm B}/B$ and ${\bm e}_{\varphi\perp}$ being vertical to ${\bm e}_{\varphi\parallel}$.
With this symmetry, $d_0({\bm k})$ has to vanish at momenta on a symmetry line ${\bm k}=k{\bm e}_{\varphi\parallel}$
so that
if the band touching happens at these momenta,
then the first-order topological phase transition arises (see the proof in Appendix C).

Specifically, for a square lattice geometry as we consider, there exists a reciprocal lattice reflection symmetry when
$\varphi=n\pi/4$ with $n$ being
an integer. At $k_F=k_{F1}$, the jump in $\sigma_H$ is associated with gap closings
at momenta on the symmetry line, as shown in Fig.~\ref{fig3}. Because of the symmetry,
the energy at the crossing points must vanish, giving rise to the
first-order topological phase transition.
Although the energy gap remains closed when we vary $\varphi$ [see Fig.~\ref{fig3}(c4)],
for other $\varphi$, such as $\varphi=0$,
$d_0$ is not enforced to vanish at momenta along $k_x=k_y$ so that the first-order phase transition does not occur [see the blue line in Fig.~\ref{fig3}(a))].
However, for $\varphi=0$, we see the occurrence of a first-order phase transition at $k_F=k_{F2}$.
There, the band touching occurs at the outer four valleys on the $k_x$ and $k_y$ axes; the band touching
on the $k_x$ axis is protected to occur at zero energy, resulting in the first-order topological quantum phase transition.
In Appendix C, we have also demonstrated the universality of the reciprocal lattice reflection symmetry protected topological phase transitions in Shiba metals, which still works in the multi-band scenario.

We also want to note that the reciprocal lattice reflection symmetry in topological Shiba metals is composed of two geometric factors, i.e., the configuration of impurity lattice and the polarization of the magnetic moments. Any disturbance on these two factors, such as structural disorder and polarization disorder, would break the protection of topological quantum phase transition.

\begin{figure}
	\includegraphics[width=2.4in]{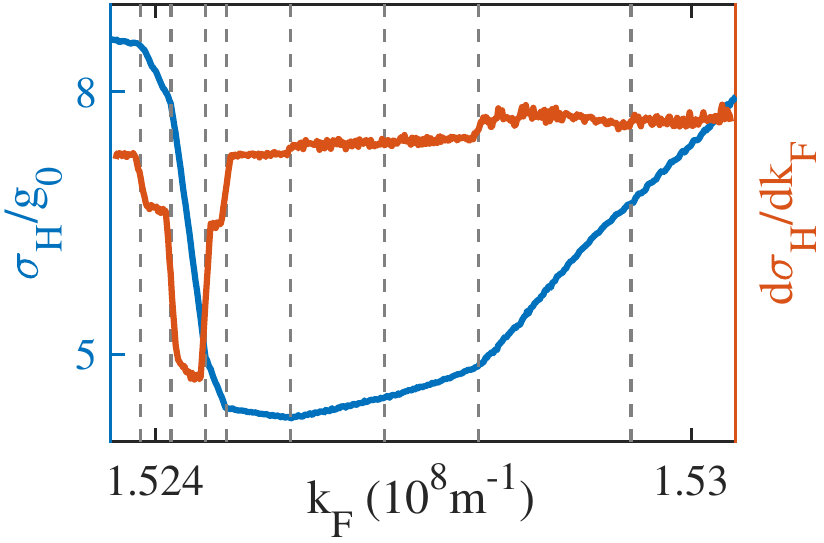}
	\caption{(Color online)
		Intrinsic thermal Hall conductance versus the Fermi wavevector $k_F$, showing many second-order topological quantum phase transitions (highlighted by dashed gray lines)
		revealed by the discontinuous change in $d\sigma_H/dk_F$.
		Since the phase transitions correspond to the Lifshitz phase transitions, we determine the transition points
		by numerically identifying the appearance or disappearance of the electron (or hole) pockets in
		the Fermi surface.
		Here, all the parameters are the same as those in Fig.~\ref{fig3}(a) except that $\varphi=\piup/8$.
	}
	\label{fig4}
\end{figure}

When $\varphi=0$, at $k_F=k_{F1}$, the first-order phase transition disappears because the gap closing points are not
pinned at zero energy [see Fig.~\ref{fig3}(c4)]. Interestingly, there appear two second-order quantum phase transitions
around this point revealed by discontinuous changes in ${\partial\sigma_H}/{\partial k_F}$.
Such a phase transition arises from the Lifshitz transition where the topology of the Fermi surface changes.
Specifically, as we increase $k_F$, the conduction band declines and the valence band rises,
so that these bands approach and then cross the zero energy,
generating an electron pocket in the conduction band and a hole pocket in the valence band [see Fig.~\ref{fig3}(c3)]
corresponding to a sharp change in the Fermi surface.
Once the pockets appear, the integral of the Berry curvature around the electron (hole) pocket
is approximated by $\Omega_2({\bm k_0})\delta S$ [$\Omega_1(-{\bm k_0})\delta S=-\Omega_2({\bm k_0})\delta S$],
where $\bm k_0$ and $\delta S$ denote the momentum and the area of the electron pocket, respectively.
The derivative of the intrinsic thermal Hall conductance contributed by the two pockets
is proportional to
$2\Omega_2({\bm k_0}) d S/d{k_F}$. Clearly, this derivative develops a discontinuous change from zero to a nonzero value
as the pockets appear, leading to a second-order quantum phase transition manifesting in
the singularity of the intrinsic thermal Hall conductance.
In fact, such second-order topological quantum phase transitions are widespread in a Shiba metal [see Fig.~\ref{fig4}]
due to the ubiquitous existence of pocket structures in the energy bands, which is attributed to the long-range hopping.
Another manifestation of the long-range hopping is the high thermal Hall conductance.
In fact, it can be much higher, but it is harder to identify the phase transitions numerically.

\section{conclusion}
In summary, we have theoretically predicted the existence of topological Shiba metals in a magnet-superconductor hybrid system
subject to a very weak in-plane magnetic field.
The topological Shiba metallic phase arises due to the formation of tilted magnetization of magnetic impurities.
Such a metallic phase exhibits intrinsic thermal Hall conductance with large nonquantized values
and undergoes many second-order quantum phase transitions for the intrinsic thermal Hall conductance.
We also find a new mechanism (reciprocal lattice reflection symmetry) that protects the
first-order topological quantum phase transitions for the intrinsic thermal Hall conductance.
Our work thus opens the door for studying topological metallic phases in
Shiba lattices.

\begin{acknowledgments}
This work is supported by the National Natural Science Foundation of China (Grant No. 11974201),
Tsinghua University Dushi Program and Shanghai Qi Zhi Institute.
\end{acknowledgments}
\begin{appendix}
\section{Magnetic order of an impurity lattice}

In this appendix, we will fill in the details about the formation of the ferromagnetic order.
The full Hamiltonian of the Shiba metal, including magnetic impurities and the underlying superconducting substrate, is presented as
\begin{eqnarray}
	H &=& H_{e0} + H_m + H_\mathrm{s-d}\label{Htotal}\\
	&=&\zeta\tau_z+\alpha_R(\vec{\sigma}\times\vec{k})_z\tau_z+\Delta\tau_x+m_\parallel\sigma_\parallel\nonumber\\
	&&+\sum_i [-\frac{D}{2}S_{i,z}^2+M_\parallel S_{i,\parallel} - J  (\vec{S}_i \cdot \vec{\sigma}) \delta(\vec{r}-\vec{r}_i) ].\nonumber
\end{eqnarray}
Here the first term,
\begin{equation}
	H_{e0} = \zeta\tau_z+\alpha_R(\vec{\sigma}\times\vec{k})_z\tau_z+\Delta\tau_x+m_\parallel\sigma_\parallel,
\end{equation}
is the Hamiltonian of electrons in $k$-space, where $\zeta=\frac{k^2}{2m}-E_F$ is the kinetic energy of free electrons with $m$ ($E_F$) denoting the effective mass (Fermi energy), $\alpha_R$ is the Rashba coefficient, $\Delta$ denotes the $s$-wave superconducting order parameter, $m_\parallel$ denotes the Zeeman splitting of electrons, $\sigma_\parallel=\vec{\sigma}\cdot\vec{e}_{\varphi\parallel}$ with $\vec{e}_{\varphi\parallel}$ being the unit vector along the direction of the magnetic field, and the Pauli matrices $\sigma$ and $\tau$ are defined on the spin and particle-hole subspaces, respectively.
The second term
\begin{equation}
	H_m = \sum_i \big(-\frac{D}{2}S_{i,z}^2+M_\parallel S_{i,\parallel}),
\end{equation}
which is the energy of the magnetic impurities, consists of the crystal field anisotropy term $-D S_{i,z}^2/2$ and the Zeeman splitting term $M_\parallel S_{i,\parallel}$ with $\vec{S}_i$ being the classical spin of the $i$th impurity atom.
Finally, the magnetic impurities and electrons are coupled via the s-d exchange interaction, which is given by
\begin{equation}
	H_\mathrm{s-d} = - J \sum_i  (\vec{S}_i \cdot \vec{\sigma}) \delta(\vec{r}-\vec{r}_i).
\end{equation}

\begin{figure}[t]
	\includegraphics[width=\linewidth]{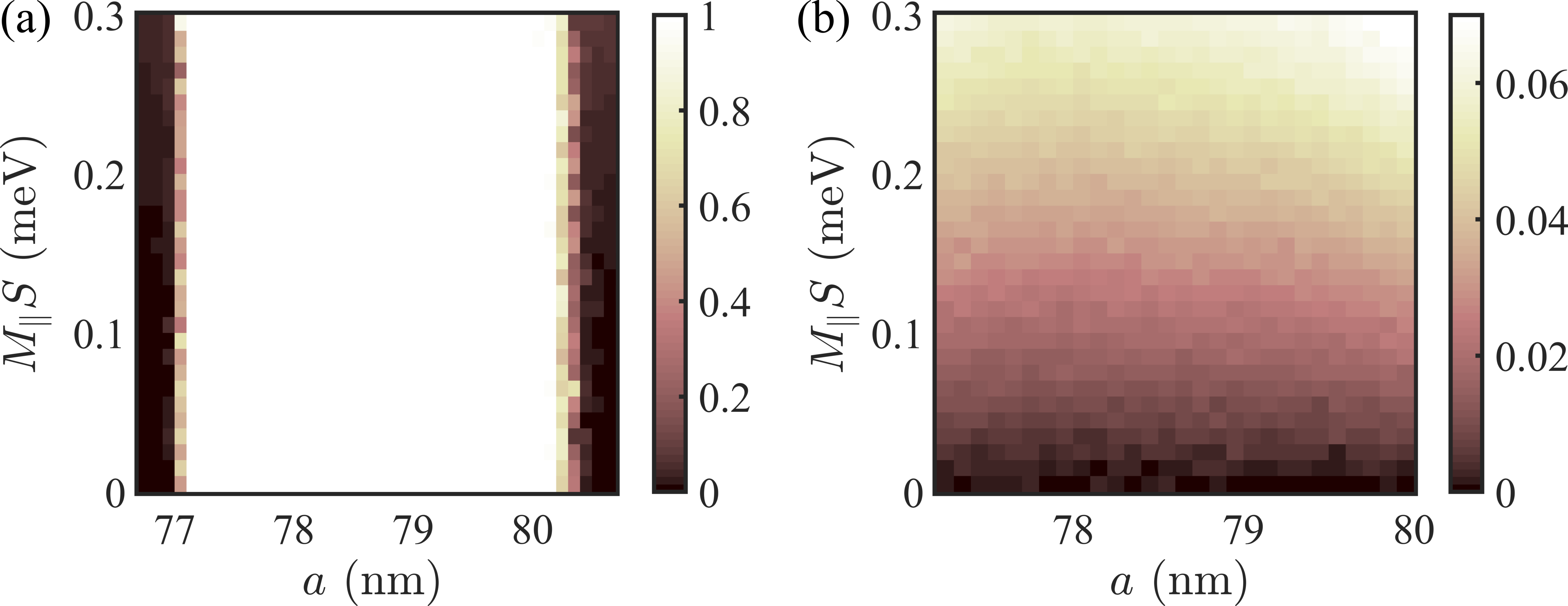}
	\caption{(a) The renormalized magnetization amplitude $\langle \vec{S}\rangle^2/S^2$ and (b) the polar angle $\theta/\pi$ as a function of the lattice constant $a$ and the Zeeman field $M_\parallel S$ for a $20\times 20$ lattice.
		Here we change the arthmuthal angle to $\varphi=\pi/4$, while other parameters are the same as those in Fig.~2 in the main text.
		The results indicate that the ferromagnetic regime still exists when $\varphi$ is varied.
	}
	\label{figS1}
\end{figure}

When only the impurities are concerned, we neglect the electron Hamiltonian $H_{e0}$ and arrive at an effective Hamiltonian describing the magnetic impurities,
\begin{equation}
	\label{HS}
	H_\mathrm{S}=\sum_i(-\frac{D}{2}S_{i,z}^2+M_\parallel S_{i,\parallel})+H_\mathrm{RKKY},
\end{equation}
where the RKKY interaction $H_\mathrm{RKKY}$ can be obtained by the second order perturbation theory \cite{Utsumi2004PRB}.
In the presence of Rashba spin-orbit coupling (SOC), the RKKY interaction takes the form of~\cite{Utsumi2004PRB,heimes2015interplay}
\begin{eqnarray}
	H_\mathrm{RKKY}&=&-m\left(\frac{Jk_F}{\pi}\right)^2\sum_{ij}\frac{\sin(2k_Fr_{ij})}{(2k_Fr_{ij})^2}\nonumber\\
	&&\{\cos(2m\alpha_Rr_{ij})\vec{S}_i\cdot\vec{S}_j\nonumber\\
	&&+[1-\cos(2m\alpha_Rr_{ij})](\vec{S}_i\cdot\vec{e}^\perp_{ij})(\vec{S}_j\cdot\vec{e}^\perp_{ij})\nonumber\\
	&&+\sin(2m\alpha_Rr_{ij})(\vec{S}_i\times\vec{S}_j)\cdot\vec{e}^\perp_{ij}\},\label{RKKY}
\end{eqnarray}
where $J$ denotes the strength of the s-d exchange coupling, $\vec{e}^\perp_{ij}=\vec{e}_z\times\vec{e}_{ij}$ and $\vec{e}_{ij}$ represents the unit vector from site $i$ to $j$.
Using $E_F = \frac{k_F^2}{2m}$ and $\alpha=\frac{mJS}{2}=\sqrt{\frac{1-\varepsilon}{1+\varepsilon}}$, Eq.~(\ref{RKKY}) can be rewritten as
\begin{eqnarray}
	H_\mathrm{RKKY}&=&-\frac{8E_F}{\piup^2}\frac{1-\varepsilon}{1+\varepsilon}\sum_{ij}\frac{\sin(2k_Fr_{ij})}{(2k_Fr_{ij})^2}\nonumber\\
	&&\left\{\cos(2m\alpha_Rr_{ij})\frac{\vec{S}_i}{S}\cdot\frac{\vec{S}_j}{S}\right.\nonumber\\
	&&+[1-\cos(2m\alpha_Rr_{ij})](\frac{\vec{S}_i}{S}\cdot\vec{e}^\perp_{ij})(\frac{\vec{S}_j}{S}\cdot\vec{e}^\perp_{ij})\nonumber\\
	&&\left.+\sin(2m\alpha_Rr_{ij})(\frac{\vec{S}_i}{S}\times\frac{\vec{S}_j}{S})\cdot\vec{e}^\perp_{ij}\right\}.\label{HRKKY}
\end{eqnarray}
Here $\varepsilon = (1-\alpha^2)/(1+\alpha^2)$ represents the position of a YSR state in the superconducting gap (detailed in the next section), and in this work we set $\varepsilon=0.2$.
Although the derivation in Ref.~\cite{Utsumi2004PRB} does not consider the superconducting term $\Delta \tau_x$ and the Zeeman term $m_\parallel \sigma_\parallel$, we note that when two adjacent impurities are not too distant from each other, i.e., $k_F r<\xi/r$, the effect of superconducting pairing $\Delta$ is negligible~\cite{yao2014enhanced}, and the electronic Zeeman term $m_\parallel$ can also be omitted since it is much smaller than $\Delta$, thus Eqs.~(\ref{RKKY}) and (\ref{HRKKY}) are still valid.
Here $\xi=v_F/\Delta$ is the superconducting coherence length, and $v_F=k_F/m$ is the Fermi velocity. In this paper we set $\xi=1200$ nm.
In the context, we use Eq.~(\ref{HS}) and (\ref{HRKKY}) to calculate the phase diagram of the impurities (Fig.~\ref{fig2} for $\varphi=0$ and Fig.~\ref{figS1} for $\varphi=\pi/4$), where the constraint $k_F r<\xi/r$ is obeyed.

Based on Eq.~(\ref{HRKKY}), we see that without Rashba SOC and superconductivity, the RKKY interaction between two nearest-neighboring impurities is proportional to the inner product of their spins
with the coefficient proportional to $-\sin(2k_F a)/(2k_F a)^2$ [see Eq.~(\ref{HRKKY})]. As a result, ferromagnetism occurs when this coefficient is negative between two adjacent impurities.
The RKKY interaction changes sign upon a change of the Shiba lattice constant by $\pi/(2k_F)$, which is about 10 nm if we take $k_F=1.57\times10^8\,m^{-1}$.
This is the reason why the ferromagnetic phase is sensitive to the Shiba lattice constant $a$.
With the SOC, other two terms arise [see Eq.~(\ref{HRKKY})]. These extra terms mitigate the ferromagnetic interaction effects and may thus reduce the ferromagnetic regime
to about 3 nm (see Fig.~\ref{fig2}).

\section{Effective Hamiltonian for Shiba states}

In this appendix, we will provide a detailed derivation of the effective tight-binding Hamiltonian for Shiba lattices.
The derivation of YSR states under a weak magnetic field is based on Ref.~\cite{Oppen PRB2013}, and the derivation of the Hamiltonian for Shiba lattices follows Refs.~\cite{Ojanen PRL2015,Ojanen nat.commun2018}.

\subsection{YSR states under a weak magnetic field}

In the ferromagnetic regime, the magnetic impurities can be treated as fixed classical spins. As a result, the Hamiltonian for electrons $H_e$ can be decoupled from the impurity Hamiltonian $H_m$, which is given by
\begin{equation}
	H_e = H^{(0)} + \Delta H + \sum_i H_i,
\end{equation}
with
\begin{eqnarray}
	H^{(0)}&=&\zeta\tau_z+\alpha_R(\vec{\sigma}\times\vec{k})_z\tau_z+\Delta\tau_x,\\
	\Delta H&=&m_\parallel\sigma_\parallel,\\
	H_i&=&-JS \sigma_i \delta(\vec{r}-\vec{r}_i).
\end{eqnarray}

We first focus on the substrate Hamiltonian $H^{(0)}$, which is in the Nambu representation with the basis being $\{ \psi_\uparrow(\vec{k}), \psi_\downarrow(\vec{k}), \psi_\downarrow^\dagger(-\vec{k}), -\psi_\uparrow^\dagger(-\vec{k}) \}$.
The Green's function of the unperturbed Hamiltonian $H^{(0)}$ in $k$-space is given by
\begin{eqnarray}
	\label{Gk}
	&&G_0(\vec{k};E)= \nonumber \\
	 &&\frac{1}{2}\sum_{\nu=\pm1}\frac{E+\zeta_\nu\tau_z+\Delta\tau_x}{E^2-\zeta_\nu^2-\Delta^2}\left[1+\nu\left(\frac{k_y}{k}\sigma_x-\frac{k_x}{k}\sigma_y\right)\right]
\end{eqnarray}
where $\zeta_\nu=\zeta+\nu\alpha_Rk$ denotes the electron energy in two spin-polarized branches.
The real-space Green's function is available by Fourier transforming Eq.~(\ref{Gk}):
\begin{eqnarray}
	G_0(\vec{r};E)&=&\frac{1}{2}\sum_{\nu=\pm1}\int\dif\vec{k}\frac{\me^{\mi\vec{k}\cdot\vec{r}}}{(2\piup)^2}\frac{E+\zeta_\nu\tau_z+\Delta\tau_x}{E^2-\zeta_\nu^2-\Delta^2}\nonumber\\
	&&\left[1+\nu\left(\frac{k_y}{k}\sigma_x-\frac{k_x}{k}\sigma_y\right)\right]\label{Gr}.
\end{eqnarray}
Specifically, the $r=0$ case $G_0(0;E)$ in the low energy regime $E\ll\Delta$ takes a spin-independent form of
\begin{equation}
	G_0(0;E) \approx -\frac{m}{2}\frac{E+\Delta\tau_x}{\sqrt{\Delta^2-E^2}}.\label{Gr0}
\end{equation}

The Green's function $G_0$ depicts the propagation of electrons in the substrate when the magnetic field is absent.
In the presence of a weak in-plane magnetic field, the corresponding Green's function can be written as $G = G_0 + \Delta G$, where the variation $\Delta G = G_0 \Delta H G_0 + G_0 \Delta H G_0 \Delta H G_0 + \cdots \approx G_0 \Delta H G_0$. In the real space, we have
\begin{equation}
	\label{DeltaG}
	\Delta G(\vec{r}; E) \approx \int\dif\vec{r}' G_0(\vec{r}-\vec{r}';E) \Delta H G_0(\vec{r}'; E).
\end{equation}
In our Shiba metal model, the in-plane magnetic field is sufficiently weak so that $m_\parallel\ll\Delta$, thus we have $\Delta H\ll H^{(0)}(\vec{r})$. The Green's function $G_0 (\vec{r};E)$ is the inverse of $E-H^{(0)}$, which gives
\begin{equation}
	G_0(\vec{r};0)=-\int\dif\vec{r}' G_0(\vec{r}-\vec{r}';0) H^{(0)}(\vec{r}') G_0(\vec{r}';0).\label{DeltaG_compare}
\end{equation}
Comparing Eq.~(\ref{DeltaG}) and (\ref{DeltaG_compare}), it is obvious that $\Delta G (\vec{r};0)$ is negligible compared with $G_0(\vec{r};0)$.

Furthermore, let's take a look at the $\vec{r}=0$ case to estimate the effect of a weak magnetic field on the Green's function:
\begin{equation}
	\Delta G (0;E) = \int\frac{\dif\vec{k}}{(2\piup)^2}G_0(\vec{k};E)\Delta HG_0(\vec{k};E).
\end{equation}
Denoting $A_\nu$ and $B_\nu$ as
\begin{eqnarray}
	A_\nu&=&\frac{E+\zeta_\nu\tau_z+\Delta\tau_x}{E^2-\zeta_\nu^2-\Delta^2},\\
	B_\nu&=&\nu\frac{E+\zeta_\nu\tau_z+\Delta\tau_x}{E^2-\zeta_\nu^2-\Delta^2}(\sin\psi_k\sigma_x-\cos\psi_k\sigma_y),
\end{eqnarray}
where $\sin\psi_k= k_y / k$ and $\cos\psi_k=k_x / k$, we have
\begin{eqnarray}
	&&\Delta G(0;E)=\int_{0}^{2\piup}\frac{\dif\psi_k}{2\piup}\int_0^\infty\frac{k\dif k}{2\piup}\\
	&&\frac{A_++A_-+B_++B_-}{2}m_\parallel \sigma_\parallel\frac{A_++A_-+B_++B_-}{2}.\nonumber
\end{eqnarray}
Crossing terms concerning $B_{\nu}$ vanish under the integral $\int\dif\psi_k\cdots$, so we arrive at
\begin{equation}
	\Delta G(0;E) = \int_{0}^{2\piup}  \frac{\dif\psi_k}{2\piup}\int_0^\infty\frac{k\dif k}{2\piup}\frac{(A_++A_-)^2}{4}m_\parallel \sigma_\parallel.
\end{equation}
Each matrix element of $A_\nu$ is a function of $k$, which reaches its peak at $k_F^\nu=k_F(\sqrt{1+\lambda^2}-\nu\lambda)$ with $\lambda=\alpha_R/v_F$ being the dimensionless Rashba coefficient. The width of this peak depends on $\Delta/E_F$, which is extremely narrow. We have numerically checked that the peaks in $A_+$ and $A_-$ are completely mismatched, which makes the crossing term $A_+A_-$ negligible.
Moreover, with $\alpha_R k_F\ll E_F$, we have $A_\nu(k)\approx A(k+\nu m\alpha_R)$ where $A$ is obtained by replacing $\zeta_\nu$ with $\zeta$ in $A_\nu$. In this context, when $E\ll\Delta$ we have
\begin{eqnarray}
	&&\Delta G(0;E)\nonumber\\
	&&\quad\approx\int_{0}^{2\piup} \frac{\dif\psi_k}{2\piup}\int_0^\infty\frac{k\dif k}{2\piup}\frac{A(k+m\alpha_R)^2+A(k-m\alpha_R)^2}{4}m_\parallel \sigma_\parallel\nonumber\\
	&&\quad\approx\int_{0}^{2\piup} \frac{\dif\psi_k}{2\piup}\int_0^\infty\frac{k\dif k}{4\piup}A(k)^2m_\parallel \sigma_\parallel\nonumber\\
	&&\quad=\int_0^\infty\frac{k\dif k}{4\piup}\left[\frac{E+\zeta\tau_z+\Delta\tau_x}{E^2-\zeta^2-\Delta^2}\right]^2m_\parallel \sigma_\parallel\nonumber\\
	&&\quad=\frac{m}{4\piup}\int_{-E_F}^\infty\dif\zeta\left[\frac{E+\zeta\tau_z+\Delta\tau_x}{E^2-\zeta^2-\Delta^2}\right]^2m_\parallel \sigma_\parallel\nonumber\\
	&&\quad\approx \frac{m}{4\piup} \int_{-\infty}^\infty \dif\zeta \left[\frac{E+\zeta\tau_z+\Delta\tau_x}{E^2-\zeta^2-\Delta^2}\right]^2m_\parallel \sigma_\parallel\nonumber\\
	&&\quad=m_\parallel m\frac{\Delta^2+E\Delta\tau_x}{4(\Delta^2-E^2)^\frac{3}{2}}\sigma_\parallel.\label{DG}
\end{eqnarray}
It is obvious from Eq.~(\ref{DG}) that near $E=0$ we have $\Delta G \sim m_\parallel / \Delta$, which is negligible.

Next we consider a single magnetic impurity on the substrate.
The impurity Hamiltonian is given by
\begin{equation}
	H_i(\vec{r})=-JS\sigma_i\delta(\vec{r}-\vec{r}_i),
\end{equation}
where $S$ is the magnitude of the classical impurity spin, $\sigma_i=\vec{n}_i\cdot\vec{\sigma}$ with $\vec{n}_i=\vec{S}_i/S$, and $\vec{r}_i$ denotes the position of the $i$th impurity atom.
The single impurity system in the real space is described by
\begin{equation}
	[H^{(0)}(\vec{r})+\Delta H+H_i(\vec{r})]\Psi(\vec{r})=E\Psi(\vec{r}).\label{single}
\end{equation}
In searching for low energy subgap YSR states, we apply the Green's function $G_0 + \Delta G = (E-H^{(0)}-\Delta H)^{-1}$ to Eq.~(\ref{single}) and get
\begin{equation}
	\label{GH0}
	[G_0(\vec{r},\vec{r}';E)+\Delta G(\vec{r},\vec{r}';E)] H_i (\vec{r})\Psi(\vec{r})=\Psi(\vec{r}) \delta(\vec{r} - \vec{r}').
\end{equation}
Integrating over $\vec{r}$ and letting $\vec{r}'=\vec{r}_i$, we obtain
\begin{equation}
	\label{GH1}
	[G_0(\vec{r}_i,\vec{r}_i;E)+\Delta G(\vec{r}_i,\vec{r}_i;E)] (-JS \sigma_i) \Psi(\vec{r}_i) = \Psi(\vec{r}_i).
\end{equation}
\begin{widetext}
Note that $G_0(\vec{r}_i, \vec{r}_i;E)=G(0;E)$ and $\Delta G(\vec{r}_i, \vec{r}_i;E) = \Delta G(0;E)$, thus by Eq.~(\ref{Gr0}) and Eq.~(\ref{DG}) we have
\begin{equation}
	\left(1-\frac{mJS}{2} \frac{E+\Delta\tau_x}{\sqrt{\Delta^2-E^2}} \sigma_i +\frac{mJS}{2}m_\parallel \frac{\Delta^2+E\Delta\tau_x}{2(\Delta^2-E^2)^\frac{3}{2}}\sigma_\parallel\sigma_i\right)\Psi(\vec{r}_i)=0.
\end{equation}
Substituting $\sigma_i=\cos\theta\sigma_z+\sin\theta\sigma_\parallel$ and denoting $\sigma_\perp=-\mi\sigma_z\sigma_\parallel$, we arrive at
\begin{equation}
	\left[\frac{2}{mJS} - \frac{\Delta}{\sqrt{\Delta^2-E^2}} \tau_x\sigma_i
	+\frac{m_\parallel}{2\Delta}  \frac{\sin\theta-\mi\cos\theta\sigma_\perp}{(1-(E/\Delta)^2)^{\frac{3}{2}}}\right]
	\Psi(\vec{r}_i)
	=\frac{E}{\sqrt{\Delta^2-E^2}}\left[\sigma_i-\frac{m_\parallel }{2\Delta}\frac{\sin\theta-\mi\cos\theta\sigma_\perp}{1-(E/\Delta)^2}\tau_x\right]\Psi(\vec{r}_i).
\end{equation}
In the $\tau_x=+1$ sector, we have
\begin{equation}
	\label{Shiba1}
	\left[\frac{2}{mJS}\sqrt{1-\frac{E^2}{\Delta^2}}-\sigma_i+\frac{m_\parallel }{2\Delta}\frac{\sin\theta-\mi\cos\theta\sigma_\perp}{1-(E/\Delta)^2}\right]\Psi(\vec{r}_i)=\frac{E}{\Delta}\left[\sigma_i-\frac{m_\parallel }{2\Delta}\frac{\sin\theta-\mi\cos\theta\sigma_\perp}{1-(E/\Delta)^2}\right]\Psi(\vec{r}_i),
\end{equation}
and in the $\tau_x=-1$ sector, we have
\begin{equation}
	\label{Shiba2}
	\left[\frac{2}{mJS}\sqrt{1-\frac{E^2}{\Delta^2}}+\sigma_i+\frac{m_\parallel }{2\Delta}\frac{\sin\theta-\mi\cos\theta\sigma_\perp}{1-(E/\Delta)^2}\right]\Psi(\vec{r}_i)=\frac{E}{\Delta}\left[\sigma_i+\frac{m_\parallel }{2\Delta}\frac{\sin\theta-\mi\cos\theta\sigma_\perp}{1-(E/\Delta)^2}\right]\Psi(\vec{r}_i).
\end{equation}
\end{widetext}
When $m_\parallel =0$, Eq.~(\ref{Shiba1}) and (\ref{Shiba2}) suggest that the in-gap YSR state at the $i$th impurity is described by $|\uparrow +_i \rangle=\psi_i(\vec{r})|\uparrow +\rangle$ and $|\downarrow -_i \rangle=\psi_i^*(\vec{r})|\downarrow -\rangle$ with energy $E_{\pm}=\pm \varepsilon\Delta$, where $\varepsilon=(1-\alpha^2)/(1+\alpha^2)$ and $\alpha = mJS/2$.
Here $\psi_i(\vec{r})$ describes the amplitude of the YSR state near the $i$th impurity in the real space, $|+\rangle$ and $|-\rangle$ denote the eigenstates of $\tau_x$ in the particle-hole subspace, and the spin polarization of the unperturbed YSR state is aligned with the magnetic impurities, with $|\uparrow\rangle$ and $|\downarrow\rangle$ being exact eigenstates of $\sigma_i$, i.e. $\sigma_i |\uparrow\rangle = |\uparrow\rangle$ and $\sigma_i |\downarrow\rangle = - |\downarrow\rangle$.
In the presence of the Zeeman perturbation, the spin parts of the YSR states become
\begin{equation}
	|\uparrow' \rangle \approx
	|\uparrow \rangle -\frac{m_\parallel \cos\theta}{4\Delta(1-\varepsilon^2)} |\downarrow \rangle
\end{equation}
and
\begin{equation}
	|\downarrow' \rangle \approx
	|\downarrow \rangle + \frac{m_\parallel \cos\theta}{4\Delta(1-\varepsilon^2)} |\uparrow \rangle,
\end{equation}
and the energy of the YSR states is given by $E_{\pm}' = \pm \varepsilon' \Delta$ with
\begin{equation}
	\varepsilon' \approx \varepsilon +
	\frac{m_\parallel \sin\theta}{2\Delta} \frac{1-\varepsilon}{1+\varepsilon}.
\end{equation}
In Cartesian coordinates, $\vec{n}_i=(\sin\theta\cos\varphi,$ $\sin\theta\sin\varphi,$ $\cos\theta)$, and the YSR state in the eigenbasis of $\sigma_z$ is
\begin{equation}
	|\uparrow' \rangle =
	\begin{pmatrix}
		\me^{-\mi\frac{\varphi}{2}}\cos\frac{\theta'}{2}	\\
		\me^{\mi\frac{\varphi}{2}}\sin\frac{\theta'}{2}
	\end{pmatrix}
	\quad
	|\downarrow' \rangle =
	\begin{pmatrix}
		-\me^{-\mi\frac{\varphi}{2}}\sin\frac{\theta'}{2}	\\
		\me^{\mi\frac{\varphi}{2}}\cos\frac{\theta'}{2}
	\end{pmatrix}\label{YSR}
\end{equation}
where $\theta'=\theta-\Delta\theta$ with the deviation $\Delta\theta \approx \frac{m_\parallel \cos\theta}{2\Delta(1-\varepsilon^2)}$, which means that the Zeeman field perturbation slightly modifies the polar angle of the polarization in YSR state by $\Delta\theta$.
Since $m_\parallel \ll\Delta$, the deviations in $\varepsilon$ and $\theta$ are both negligible.

\subsection{Effective Hamiltonian for Shiba lattices}
In a Shiba lattice which consists of multiple magnetic impurities, the YSR states at different sites are coupled with each other in the superconducting substrate.
This coupling process is governed by $G(\vec{r}\neq0)$.
For this reason, the first goal in this section is to obtain a numerically computable form of Eq.~(\ref{Gr}), which can be divided into two branches $\nu=\pm1$.
Considering the SOC modified free electron energy $\zeta_\nu$, the corresponding modified Fermi wavevector is $k_F^\nu$.
In each branch, the integral is mainly contributed by the regions where $\zeta_\nu(k)\sim0$.
We can linearize $\zeta_\nu(k)$ near the Fermi surface $\zeta_\nu(k_F^\nu)=0$, yielding $\zeta_\nu(k)=v_F^\nu(k-k_F^\nu)$, where $v_F^\nu=\sqrt{1+\lambda^2}v_F$.
\begin{widetext}
The Green's function Eq.~(\ref{Gr}) can be put as
\begin{equation}
	G_0(\vec{r};E)=\frac{m}{2}\sum_{\nu=\pm1}\left(1-\nu\frac{\lambda}{\sqrt{1+\lambda^2}}\right)\int\dif\psi_k\dif\zeta\frac{\me^{\mi (k_F^\nu+\frac{\zeta}{v_F})r\cos(\psi_k-\psi_r)}}{(2\piup)^2}\frac{E+\zeta\tau_z+\Delta\tau_x}{E^2-\zeta^2-\Delta^2}[1+\nu(\sin\psi_k\sigma_x-\cos\psi_k\sigma_y)],
\end{equation}
where $\psi_k$ and $\psi_r$ are polar angles of $\vec{k}$ and $\vec{r}$, respectively.
Substituting $\phi=\psi_k-\psi_r$, and neutralizing the odd part with respect to $\phi$, we have
\begin{equation}
	G_0(\vec{r};E)=\frac{m}{2}\sum_{\nu=\pm1}\left(1-\nu\frac{\lambda}{\sqrt{1+\lambda^2}}\right)\int\dif\phi\dif\zeta\frac{\me^{\mi (k_F^\nu+\frac{\zeta}{v_F})r\cos\phi}}{(2\piup)^2}\frac{E+\zeta\tau_z+\Delta\tau_x}{E^2-\zeta^2-\Delta^2}[1+\nu\cos\phi(\sin\psi_r\sigma_x-\cos\psi_r\sigma_y)],
\end{equation}
The Green's function in this form can be expressed via Bessel functions using the following identity relations:
\begin{eqnarray}
	\int_{-\infty}^\infty\frac{\dif\zeta}{\piup}\int_0^{2\piup}\frac{\dif\phi}{2\piup}\frac{\zeta\me^{\mi (k_F+\frac{\zeta}{v_F})r\cos\phi}\cos\phi}{E^2-\zeta^2-\Delta^2}&=&-\mi\mathrm{Re}\left[\mi J_1(k_Fr+\mi\frac{r}{\xi_E})+\frac{2}{\piup}-H_1(k_Fr+\mi\frac{r}{\xi_E})\right]\label{ir1}, \\
	\int_{-\infty}^\infty\frac{\dif\zeta}{\piup}\int_0^{2\piup}\frac{\dif\phi}{2\piup}\frac{\me^{\mi (k_F+\frac{\zeta}{v_F})r\cos\phi}\cos\phi}{E^2-\zeta^2-\Delta^2}&=&\frac{-\mi}{\sqrt{\Delta^2-E^2}}\mathrm{Im}\left[\mi J_1(k_Fr+\mi\frac{r}{\xi_E})+\frac{2}{\piup}-H_1(k_Fr+\mi\frac{r}{\xi_E})\right]\label{ir2}, \\
	\int_{-\infty}^\infty\frac{\dif\zeta}{\piup}\int_0^{2\piup}\frac{\dif\phi}{2\piup}\frac{\zeta\me^{\mi (k_F+\frac{\zeta}{v_F})r\cos\phi}}{E^2-\zeta^2-\Delta^2}&=&\mathrm{Im}\left[J_0(k_Fr+\mi\frac{r}{\xi_E})+\mi H_0(k_Fr+\mi\frac{r}{\xi_E})\right]\label{ir3}, \\
	\int_{-\infty}^\infty\frac{\dif\zeta}{\piup}\int_0^{2\piup}\frac{\dif\phi}{2\piup}\frac{\me^{\mi (k_F+\frac{\zeta}{v_F})r\cos\phi}}{E^2-\zeta^2-\Delta^2}&=&\frac{-1}{\sqrt{\Delta^2-E^2}}\mathrm{Re}\left[J_0(k_Fr+\mi\frac{r}{\xi_E})+\mi H_0(k_Fr+\mi\frac{r}{\xi_E})\right].\label{ir4}
\end{eqnarray}
With the help of Eq.~(\ref{ir1})--(\ref{ir4}), we can rewrite $G_0(\vec{r};E)$ into a compact form:
\begin{equation}
	G_0(\vec{r};E)=-\frac{m}{4}\left[\frac{E+\Delta\tau_x}{\sqrt{\Delta^2-E^2}}\mathrm{Re}S(r)-\tau_z\mathrm{Im}S(r)+\mi \left(\tau_z\mathrm{Re}A(r)+\frac{E+\Delta\tau_x}{\sqrt{\Delta^2-E^2}}\mathrm{Im}A(r)\right)(\sin\psi_r\sigma_x-\cos\psi_r\sigma_y)\right],
\end{equation}
where
\begin{eqnarray}
	S(r)&=&\sum_{\nu=\pm1}\left(1-\nu\frac{\lambda}{\sqrt{1+\lambda^2}}\right)\left[J_0(k_F^\nu r+\mi\frac{r}{\xi_E})+\mi H_0(k_F^\nu r+\frac{r}{\xi_E})\right], \\
	A(r)&=&\sum_{\nu=\pm1}\nu\left(1-\nu\frac{\lambda}{\sqrt{1+\lambda^2}}\right)\left[\mi J_1(k_F^\nu r+\mi\frac{r}{\xi_E})+\frac{2}{\piup}-H_1(k_F^\nu r+\mi\frac{r}{\xi_E})\right].
\end{eqnarray}
Here $J_n$ and $H_n$ are the $n$th order Bessel and Struve functions, respectively, and $\xi_E=\frac{v_F}{\sqrt{\Delta^2-E^2}}$ corresponds to the superconducting coherence length. Since we are dealing with low-energy YSR states, we can let $E=0$ and replace $\xi_E$ by $\xi$.
\end{widetext}
As we have already acquired the Green's function $G_0(\vec{r};E)$, now we are able to handle the multi-impurity system, which is described by
\begin{equation}
	\label{muti0}
	(H^{(0)}+\Delta H+\sum_iH_i)|\Psi\rangle=E|\Psi\rangle,
\end{equation}
with $H_i$ representing the Hamiltonian for the $i$th impurity. Using the Green's function for $H^{(0)}+\Delta H$ which is denoted by $G$, Eq.~(\ref{muti0}) can be derived to
\begin{equation}
	G\sum_iH_i|\Psi\rangle=|\Psi\rangle\label{muti1}.
\end{equation}
Since the perturbation of YSR states by the magnetic field is insignificant, especially in the small $\theta$ case, we can adopt the unperturbed YSR states as the complete orthogonal basis, and write the wavefunction $| \Psi \rangle$ as
\begin{equation}
	\label{basis}
	|\Psi\rangle=\frac{1}{\sqrt{N}}\sum_i|\Psi_i\rangle,
\end{equation}
where $N$ is the normalization factor and $| \Psi_i\rangle=a_i |\uparrow +_i \rangle + b_i |\downarrow -_i \rangle$ is the wave function on the $i$th impurity.
By Eq.~(\ref{muti1}) and Eq.~(\ref{basis}), we obtain
\begin{equation}
	(1-G H_i)|\Psi_i\rangle=\sum_{j\neq i}GH_j|\Psi_j\rangle.\label{muti2}
\end{equation}
\begin{widetext}
By projecting Eq.~(\ref{muti2}) on $\delta(\vec{r}-\vec{r}_i)|\uparrow + \rangle$ and $\delta(\vec{r}-\vec{r}_i)|\downarrow - \rangle$ respectively, Eq.~(\ref{muti2}) can be expressed in a matrix form:
\begin{equation}
	\label{effHam}
	\begin{aligned}
		&\begin{bmatrix}
			-\langle\uparrow + |1-G(0;E)H_\mathrm{imp}|\uparrow +\rangle	& -\langle\uparrow +|1-G(0;E)H_\mathrm{imp}|\downarrow -\rangle \\
			\langle\downarrow -|1-G(0;E)H_\mathrm{imp}|\uparrow + \rangle	& \langle\downarrow -|1-G(0;E)H_\mathrm{imp}|\downarrow -\rangle
		\end{bmatrix}
		\begin{pmatrix}
			a_i	\\
			b_i	
		\end{pmatrix}
		\\&\qquad=
		\sum_{j\neq i}
		\begin{bmatrix}
			-\langle\uparrow +|G(\vec{r}_{ij};E)H_\mathrm{imp}|\uparrow +\rangle	& -\langle\uparrow +|G(\vec{r}_{ij};E)H_\mathrm{imp}|\downarrow -\rangle \\
			\langle\downarrow -|G(\vec{r}_{ij};E)H_\mathrm{imp}|\uparrow +\rangle	& \langle\downarrow -|G(\vec{r}_{ij};E)H_\mathrm{imp}|\downarrow -\rangle
		\end{bmatrix}
		\begin{pmatrix}
			a_j	\\
			b_j	
		\end{pmatrix}
	\end{aligned}
\end{equation}
where $H_\mathrm{imp} = -JS \sigma_i$. The matrix elements on the left hand side are related to $G(0;E)$, given by
\begin{eqnarray}
	\langle\uparrow + | 1-G(0;E) H_\mathrm{imp} | \uparrow +\rangle &=&\frac{JSm}{2}\frac{\varepsilon\Delta-E}{\Delta},\\
	\langle\downarrow - | 1-G(0;E) H_\mathrm{imp} | \downarrow - \rangle &=&\frac{JSm}{2}\frac{\varepsilon\Delta+E}{\Delta}, \\
	\langle\uparrow + | 1-G(0;E) H_\mathrm{imp} | \downarrow - \rangle = \langle\downarrow - | 1&-&G(0;E) H_\mathrm{imp} | \uparrow + \rangle  =0.
\end{eqnarray}
Then we can rewrite Eq.~(\ref{effHam}) into a time-independent Schr\"odinger-like equation
\begin{equation}
	E
	\begin{pmatrix}
		a_i	\\
		b_i	
	\end{pmatrix}
	=\Delta\begin{bmatrix}
		\varepsilon	& \\
		& -\varepsilon
	\end{bmatrix}\begin{pmatrix}
		a_i	\\
		b_i	
	\end{pmatrix}+\frac{2\Delta}{JSm}\sum_{j\neq i}\begin{bmatrix}
		-\langle\uparrow +|G(\vec{r}_{ij};E)H_\mathrm{imp}|\uparrow +\rangle	& -\langle\uparrow +|G(\vec{r}_{ij};E)H_\mathrm{imp}|\downarrow -\rangle \\
		\langle\downarrow -|G(\vec{r}_{ij};E)H_\mathrm{imp}|\uparrow +\rangle	& \langle\downarrow -|G(\vec{r}_{ij};E)H_\mathrm{imp}|\downarrow -\rangle
	\end{bmatrix}\begin{pmatrix}
		a_j	\\
		b_j	
	\end{pmatrix}\label{Schrlike}.
\end{equation}

The matrix elements on the right hand side are related to $G(\vec{r}\neq0,E)$. Since the coupling between YSR states is weak, this term can be treated perturbatively so that $G(\vec{r}\neq0;E)\approx G(\vec{r}\neq0;0)$ in the low-energy regime. Using Eq.~(\ref{YSR}) and $G\approx G_0$ (since $\Delta G$ is negligible compared with $G_0$), we have
\begin{eqnarray}
	\langle\uparrow +|G(\vec{r}_{ij};0)H_\mathrm{imp}|\uparrow  + \rangle&=&\frac{mJS}{4}[\mathrm{Re}S(r_{ij})+\mi\,\mathrm{Im}A(r_{ij}) \sin\theta\sin(\psi_r-\varphi)], \\
	\langle\downarrow -|G(\vec{r}_{ij};0)H_\mathrm{imp}|\downarrow -\rangle&=&\frac{mJS}{4}[\mathrm{Re}S(r_{ij})-\mi\,\mathrm{Im}A(r_{ij})\sin\theta\sin(\psi_r-\varphi)],\\
	\langle\uparrow +|G(\vec{r}_{ij};0)H_\mathrm{imp}|\downarrow - \rangle&=&\mi\frac{mJS}{4}\mathrm{Re}A(r_{ij})[\cos\theta\sin(\psi_r-\varphi)+\mi\cos(\psi_r-\varphi)],\\
	\langle\downarrow -|G(\vec{r}_{ij};0)H_\mathrm{imp}|\uparrow +\rangle&=&-\mi\frac{mJS}{4}\mathrm{Re}A(r_{ij})[\cos\theta\sin(\psi_r-\varphi)-\mi\cos(\psi_r-\varphi)].
\end{eqnarray}
\end{widetext}
Then the effective Schr\"odinger equation can be written as
\begin{equation}
	E\Psi_i=\sum_{j}[d_0(\vec{r}_{ij})+\vec{d}(\vec{r}_{ij})\cdot\vec{\sigma}]\Psi_j
\end{equation}
where $\sigma$ denotes $2\times2$ Pauli matrix, $\Psi_i=(a_i, b_i)^T$, for $r\neq0$,
\begin{eqnarray}
	d_0(\vec{r})&=&-\mi\frac{\Delta}{2} \mathrm{Im}A(r) \sin\theta\sin(\psi_r-\varphi),\\
	d_x(\vec{r})&=&-\mi\frac{\Delta}{2} \mathrm{Re}A(r) \cos\theta\sin(\psi_r-\varphi),\\
	d_y(\vec{r})&=&\mi\frac{\Delta}{2} \mathrm{Re}A(r) \cos(\psi_r-\varphi),\\
	d_z(\vec{r})&=&-\frac{\Delta}{2} \mathrm{Re}S(r),
\end{eqnarray}
and for $r=0$,
\begin{equation}
	d_0(0)=d_x(0)=d_y(0)=0,\quad d_z(0)=\varepsilon\Delta.
\end{equation}
The k-space Hamiltonian for a square Shiba lattice is given by
\begin{eqnarray}
	H(\vec{k})&=&d_0(\vec{k})+\bm{d}(\vec{k})\cdot\bm{\sigma}\label{Hamk}, \\
	d_n(\vec{k})&=&\sum_{\vec{R}}\me^{-\mi\vec{k}\cdot\vec{R}}d_n(\vec{R}), \label{Fourier}
\end{eqnarray}
which constitute the foundation for investigation in Shiba metals.

\section{Reciprocal lattice reflection symmetry}
In this appendix, we will demonstrate how a reciprocal lattice reflection symmetry protects a first-order topological phase transition and show that such phase transitions are widespread in Shiba metals.

The first-order topological phase transitions are protected on a continuous set of points in the Brillouin zone, on which the deformation term $d_0(\vec{k})$ is enforced to vanish. Specifically, the Schr\"odinger-like Eq.~(\ref{Schrlike}) gives
\begin{eqnarray}
	d_0(\vec{k})&=&-\frac{\Delta}{JSm}\sum_{\vec{R}}\me^{-\mi\vec{k}\cdot\vec{R}}[\langle\uparrow+|G(\vec{R};0)H_\mathrm{imp}|\uparrow+\rangle\nonumber\\
	&&-\langle\downarrow-|G(\vec{R};0)H_\mathrm{imp}|\downarrow-\rangle],
\end{eqnarray}
where $\vec{R}$ runs over all the coordinates of impurities.
Using
\begin{equation}
	G(\vec{R};0)=\int\frac{\dif\vec{k}}{(2\piup)^2}\me^{\mi\vec{k}\cdot\vec{R}}G(\vec{k};0),
\end{equation}
we arrive at
\begin{eqnarray}
	d_0(\vec{k})&=&-\frac{\Delta}{JSm}\sum_{\vec{K}}[\langle\uparrow+|G(\vec{k}+\vec{K};0)H_\mathrm{imp}|\uparrow+\rangle\nonumber\\
	&&-\langle\downarrow-|G(\vec{k}+\vec{K};0)H_\mathrm{imp}|\downarrow-\rangle],\label{d0Eq111}
\end{eqnarray}
where $\vec{K}$ runs over all reciprocal lattice vectors.
Since $\vec{e}_{\varphi\perp}=\vec{e}_z\times\vec{e}_{\varphi\parallel}$ is perpendicular to the plane spanned by $\vec{e}_z$ and $\vec{e}_{\varphi\parallel}$, and $\sigma_\perp=\vec{\sigma}\cdot\vec{e}_{\varphi\perp}$, we have
\begin{equation}
	\tau_y\sigma_\perp|+\uparrow\rangle=|-\downarrow\rangle.\label{sym}
\end{equation}
At $E=0$ we have $G(\vec{k};0)=-H(\vec{k})^{-1}$. So we get
\begin{eqnarray}
	\langle\uparrow+|G(\vec{k};0)\sigma_i|\uparrow+\rangle&=&-\langle\downarrow-|\tau_y\sigma_\perp H(\vec{k})^{-1}\sigma_i\tau_y\sigma_\perp|\downarrow-\rangle\nonumber\\\\
	\langle\downarrow-|G(\vec{k};0)\sigma_i|\downarrow-\rangle&=&-\langle\downarrow-|H(\vec{k})^{-1}\sigma_i|\downarrow-\rangle,
\end{eqnarray}
based on which we reduce Eq.~(\ref{d0Eq111}) to
\begin{eqnarray}
	d_0(\vec{k})&=&\frac{\Delta}{m}\sum_{\vec{K}}\langle\downarrow-|\tau_y\sigma_\perp H(\vec{k}+\vec{K})^{-1}\sigma_i\tau_y\sigma_\perp\nonumber\\
	&&-H(\vec{k}+\vec{K})^{-1}\sigma_i|\downarrow-\rangle\label{d0k}.
\end{eqnarray}
In addition, the substrate Hamiltonian $H=H^{(0)}+\Delta H$ in $k$-space can be put as:
\begin{equation}
	H(\vec{k})=\zeta(k)\tau_z+\alpha_R(k_{\varphi\perp}\sigma_\parallel-k_{\varphi\parallel}\sigma_\perp)\tau_z+\Delta\tau_x+m_\parallel \sigma_\parallel,
\end{equation}
where $k_{\varphi\parallel}=\vec{k}\cdot\vec{e}_{\varphi\parallel}$ and $k_{\varphi\perp}=\vec{k}\cdot\vec{e}_{\varphi\perp}$. With the help of the following identities:
\begin{eqnarray}
	\sigma_\perp(\sigma_i\sigma_\perp)\sigma_\perp=-\sigma_i\sigma_\perp\\
	\sigma_\perp(\sigma_i\sigma_\parallel)\sigma_\perp=\sigma_i\sigma_\parallel,
\end{eqnarray}
we obtain
\begin{eqnarray}
	\tau_y\sigma_\perp \sigma_iH(\vec{k})\tau_y\sigma_\perp&=&\zeta(k)\sigma_i\tau_z-\alpha_R\sigma_i(k_{\varphi\perp}\sigma_\parallel+k_{\varphi\parallel}\sigma_\perp)\tau_z\nonumber\\
	&&+\Delta\sigma_i\tau_x+m_\parallel \sigma_i\sigma_\parallel\\
	\sigma_iH(k_{\varphi\parallel},k_{\varphi\perp})&=&\tau_y\sigma_\perp\sigma_iH(k_{\varphi\parallel},-k_{\varphi\perp})\tau_y\sigma_\perp\\
	H^{-1}(k_{\varphi\parallel},k_{\varphi\perp})\sigma_i&=&\tau_y\sigma_\perp H^{-1}(k_{\varphi\parallel},-k_{\varphi\perp})\sigma_i\tau_y\sigma_\perp. \label{result}
\end{eqnarray}
Substituting Eq.~(\ref{result}) into Eq.~(\ref{d0k}), we have
\begin{eqnarray}
	\label{d0}
	&&d_0(\vec{k})=  \\
	&&\frac{\Delta}{m}\sum_{\vec{K}}\langle\downarrow-|[H(\mathcal{M}(\vec{k}+\vec{K}))^{-1}-H(\vec{k}+\vec{K})^{-1}]\sigma_i|\downarrow-\rangle \nonumber
\end{eqnarray}
where $\mathcal{M}$ is the reflection operator satisfying $\mathcal{M}(k_{\varphi\parallel},k_{\varphi\perp})=(k_{\varphi\parallel},-k_{\varphi\perp})$.
Eq.~(\ref{d0}) tells us that when reciprocal lattice vectors $\vec{K}$ are symmetrically distributed beside the magnetic direction $\vec{e}_{\varphi\parallel}$, $d_0(k_{\varphi\parallel},0)$ is protected to be 0.
Because in this case we have $\mathcal{M}\vec{k}=\vec{k}$ and $\sum_{\vec{K}}=\sum_{\mathcal{M}\vec{K}}$.
For square lattice, this confinement yields $\varphi={n \piup}/{4}$ ($n=0,1,2,3$).
It is worth mentioning that although we have neglected the magnetic field disturbance on our effective Shiba lattice Hamiltonian [Eq.~(\ref{Hamk})], the above derivation is still valid in the presence of the magnetic field.

Next, we demonstrate that such protected phase transitions are widespread in Shiba metals.
To be specific, the Shiba metal is characterized by a series of system parameters.
Excluding the controlled variable $k_F$, these parameters can be grouped as a vector $\vec{P}=\{a,\lambda,\xi,\varepsilon,\cdots\}$.
At some parameter points $\vec{P}_\mathrm{RS}$, tuning $k_F$ leads to gap closing on the symmetry line ${\bm k}=k{\bm e}_{\varphi\parallel}$
in the Brillouin zone,
which is parallel to the magnetic field.
In the following, we show that all these points $\vec{P}_\mathrm{RS}$ constitute some continuous regions with the same dimension as the parameter space.

At the band touching points $\vec{k}=(k_x,k_y)$, $d_x(\vec{k})=d_y(\vec{k})=d_z(\vec{k})=0$. For square lattices with the Zeeman perturbation
omitted, the three constraints are expressed as
\begin{widetext}
\begin{eqnarray}
	&\sum_{x,y}[\sin(k_xx)\cos(k_yy)\cos\psi_r\sin\varphi-\cos(k_xx)\sin(k_yy)\sin\psi_r\cos\varphi]\mathrm{Re}A(r)=0&,\label{confine1}\\
	&\sum_{x,y}[\sin(k_xx)\cos(k_yy)\cos\psi_r\cos\varphi+\cos(k_xx)\sin(k_yy)\sin\psi_r\sin\varphi]\mathrm{Re}A(r)=0&,\label{confine2}\\
	&\varepsilon-\frac{1}{2}\sum_{x,y}\cos(k_xx)\cos(k_yy)\mathrm{Re}S(r)=0.&\label{confine3}
\end{eqnarray}
\end{widetext}
Generally, in pursuit of a band touching point on a particular trajectory $[k_x(t_k),k_y(t_k)]$ while tuning the controlled parameter $k_F$, the three constraint Eq.~(\ref{confine1})--(\ref{confine3}) must be met simultaneously with only two tunable parameters $(t_k,k_F)$, which is hard to achieve.
However, for the trajectory $[t_k\cos{n\piup}/{4},t_k\sin{n\piup}/{4}]$ with $\varphi={n\piup}/{4}$, the first constraint Eq.~(\ref{confine1}) is always satisfied, leaving only two independent constraints, corresponding to two curves in the $(t_k,k_F)$ parameter plane.
These two curves are sensitive to $\vec{P}$ and thus intersect frequently in the parameter space.
Moreover, when the two curves intersect at a certain $\vec{P}_\mathrm{RS}$, they must keep intersecting at  $\vec{P}_\mathrm{RS}+\dif\vec{P}$ where $\vec{P}_\mathrm{RS}$ is shifted by a small value.
For this reason, there is a continuous region of parameters in which band touching happens at some momenta $\vec{k}$ satisfying $\psi_k={n\piup}/{4}$ while tuning $k_F$.

In fact, the elimination of the first constraint Eq.~(\ref{confine1}) is rooted in the reflection symmetry, which is still valid in the presence of the Zeeman field perturbation.
Specifically, when the energy gap closes at $\vec{k}$,
the off-diagonal elements in Eq.~(\ref{Schrlike}) satisfy the equation
\begin{equation}
	\sum_{\vec{R}}\me^{-\mi\vec{k}\cdot\vec{R}}\langle\uparrow+|G(\vec{R};0)H_\mathrm{imp}|\downarrow-\rangle=0,
\end{equation}
which is equivalent to
\begin{equation}
	\sum_{\vec{K}}\langle\uparrow+|G(\vec{k}+\vec{K};0)H_\mathrm{imp}|\downarrow-\rangle=0.
\end{equation}
This equation imposes two constraints corresponding to its real and imaginary parts.
However, with aid of Eq.~(\ref{result}), we find that
\begin{eqnarray}
	&&\sum_{\vec{K}}\langle\uparrow+|G(\vec{k}+\vec{K};0)H_\mathrm{imp}|\downarrow-\rangle\nonumber\\
	&=&-\sum_{\vec{K}}\langle\uparrow+|G(\mathcal{M}(\vec{k}+\vec{K});0)H_\mathrm{imp}|\downarrow-\rangle^*\label{result1}.
\end{eqnarray}
With the reciprocal lattice reflection symmetry such that $\mathcal{M}\{\vec{K}\}=\{\vec{K}\}$,
when $\vec{k}$ lies on the symmetry line, Eq.~(\ref{result1}) leads to
\begin{equation}
	\mathrm{Re}\sum_{\vec{K}}\langle\uparrow+|G(\vec{k}+\vec{K};0)H_\mathrm{imp}|\downarrow-\rangle=0,
\end{equation}
which eliminates one constraint.

Previously, we consider the impurity Hamiltonian taking the form of $H_\mathrm{imp}=J\delta(\vec{r})\vec{\sigma}$. There, each impurity provides only one scattering channel corresponding to zero angular momentum and binds only one YSR state. If we put the impurity scattering term in a more general form $J(\vec{r})\vec{\sigma}$, there would be multiple channels\cite{Lutchyn PRB2016}, denoted by quantum number $l$. Each channel binds one YSR state, which is represented by two basis $|l,s\rangle$ and $|l,\bar{s}\rangle$. Here $s$ is one of $|\uparrow+\rangle$ and $|\downarrow-\rangle$; $\bar{s}$ is the other one.  Suppose there are $n$ channels in total, then the hopping matrix in the tight-binding Hamiltonian is $2n\times2n$, and the matrix elements can be denoted as $H_{l_1s_1l_2s_2}(\vec{r})$.

In this context, Eq.(\ref{d0}) and Eq.(\ref{result1}) turns into
\begin{widetext}
\begin{equation}
	H_{l_1s_1l_2s_2}(\vec{k})+H_{l_1\bar{s}_1l_2\bar{s}_2}(\vec{k})\sim\sum_{\vec{K}}\langle l_1\bar{s}_1|[H(\vec{k}+\vec{K})^{-1}-H(\mathcal{M}(\vec{k}+\vec{K}))^{-1}]\sigma_i|l_2\bar{s}_2\rangle.
\end{equation}
\end{widetext}
This term vanishes when $\vec{k}$ is aligned with both the magnetic field and a symmetry line in the reciprocal lattice. In this case, the Hamiltonian takes the form of
\begin{equation}
	H(\vec{k})=H_l^{(y)}(\vec{k})\otimes\sigma_y+H_l^{(z)}(\vec{k})\otimes\sigma_z.
\end{equation}
Here $H_l^{(y)/(z)}$ is an $n\times n$ matrix. Every eigenstate $\psi$ corresponds to another eigenstate $\sigma_x\psi$ with opposite eigenvalue. For this reason, the reciprocal lattice reflection symmetry still protects quantum phase transitions in the multi-band case.
\end{appendix}


\begin{thebibliography}{99}

\bibitem{Green PRB2000}
N. Read and D. Green,
Phys. Rev. B \textbf{61}, 10267 (2000).

\bibitem{Kitaev Phys.Usp.2001}
A. Y. Kitaev,
Phys. Usp. \textbf{44}, 131 (2001).

\bibitem{Ludwig PRB2008}
A. P. Schnyder, S. Ryu, A. Furusaki, and A. W. W. Ludwig,
Phys. Rev. B \textbf{78}, 195125 (2008).

\bibitem{Kitaev 2009}
A. Kitaev,
AIP Conf. Proc. \textbf{1134}, 22 (2009).

\bibitem{DasSarma PRL2010}
J. D. Sau, R. M. Lutchyn, S. Tewari, and S. Das Sarma,
Phys. Rev. Lett. \textbf{104}, 040502 (2010).

\bibitem{Alicea Rep.Prog.Phys.2012}
J. Alicea,
Rep. Prog. Phys. \textbf{75}, 076501 (2012).

\bibitem{Beenakker 2013}
C. W. J. Beenakker,
Annu. Rev. Condens. Matter Phys. \textbf{4} 113 (2013).

\bibitem{Ando Rep.Prog.Phys.2017}
M. Sato and Y. Ando,
Rep. Prog. Phys. \textbf{80}, 076501 (2017).

\bibitem{Wendin Rep.Prog.Phys.2017}
G. Wendin,
Rep. Prog. Phys. \textbf{80}, 106001 (2017).

\bibitem{Ferrini PRL2020}
T. Hillmann, F. Quijandr\'{i}a, G. Johansson, A. Ferraro, S. Gasparinetti, and G. Ferrini,
Phys. Rev. Lett. \textbf{125}, 160501 (2020).

\bibitem{Oppen PRB2013}
F. Pientka, L. I. Glazman, and F. von Oppen,
Phys. Rev. B \textbf{88}, 155420 (2013).

\bibitem{Ojanen PRB2014}
K. P\"{o}yh\"{o}nen, A. Weststr\"{o}m, J. R\"{o}ntynen, and T. Ojanen,
Phys. Rev. B \textbf{89}, 115109 (2014).

\bibitem{Sau PRB2015}
P. M. R. Brydon, S. Das Sarma, H.-Y. Hui, and J. D. Sau,
Phys. Rev. B \textbf{91}, 064505 (2015).

\bibitem{Ojanen PRL2015}
J. R{\"o}ntynen and T. Ojanen,
Phys. Rev. Lett. \textbf{114}, 236803 (2015).

\bibitem{Ojanen PRB2016}
J. R\"{o}ntynen and T. Ojanen,
Phys. Rev. B \textbf{93}, 094521 (2016).

\bibitem{schecter2016self}
M. Schecter, K. Flensberg, M. H. Christensen, B. M. Andersen, and J. Paaske,
Phys. Rev. B \textbf{93}, 140503(R) (2016).

\bibitem{li2016two}
J. Li, T. Neupert, Z.-J Wang, A. H. MacDonald, A. Yazdani and B. A. Bernevig, Nat. Commun. \textbf{7}, 12297 (2016).

\bibitem{Ojanen nat.commun2018}
K. P{\"o}yh{\"o}nen, I. Sahlberg, A. Weststr{\"o}m, and T. Ojanen,
Nat. Commun. \textbf{9}, 2103 (2018).

\bibitem{Shiba 1968}
H. Shiba,
Prog. Theor. Phys. \textbf{40}, 435 (1968).

\bibitem{Zhu Rev.Mod.Phys.2006}
A. V. Balatsky, I. Vekhter, and J.-X. Zhu,
Rev. Mod. Phys. \textbf{78}, 373 (2006).

\bibitem{yao2014enhanced}
N. Y. Yao, L. I. Glazman, E. A. Demler, M. D. Lukin, and J. D. Sau,
Phys. Rev. Lett. \textbf{113}, 087202 (2014).

\bibitem{refereeAadded}
R. $\check{\mathrm{Z}}$itko, J. S. Lim,R. L\'opez, and R. Aguado, Phys. Rev. B \textbf{91}, 045441 (2015).

\bibitem{hatter2015magnetic}
N. Hatter, B. W. Heinrich, M. Ruby, J. I. Pascual, and K. J. Franke,
Nat. Commun. \textbf{6} 8988 (2015).

\bibitem{ruby2016orbital}
M. Ruby, Y. Peng, F. von Oppen, B. W. Heinrich, and K. J. Franke,
Phys. Rev. Lett. \textbf{117}, 186801 (2016).

\bibitem{yang2020observation}
X. Yang, Y. Yuan, Y. Peng, E. Minamitani, L. Peng, J. -J. Xian, W. -H. Zhang, Y. -S. Fu,
Nanoscale \textbf{12}, 8174 (2020).

\bibitem{ding2021tuning}
H. Ding, Y. Hu, M. T. Randeria, S. Hoffman, O. Deb, J. Klinovaja, D. Loss, and A. Yazdani,
Proc. Natl. Acad. Sci. USA \textbf{118}, e2024837118 (2021).

\bibitem{beck2021spin}
P. Beck, L. Schneider, L. R\'{o}zsa, K. Palot\'{a}s, A. L\'{a}szl\'{o}ffy, L. Szunyogh, J. Wiebe, and R. Wiesendanger,
Nat. Commun. \textbf{12}, 2040 (2021).

\bibitem{wang2021PRL}
D. -F. Wang, J. Wiebe, R. -D. Zhong, G. -D. Gu, and R. Wiesendanger,
Phys. Rev. Lett. \textbf{126}, 076802 (2021).

\bibitem{Wiebe2021NP}
L. Schneider, P. Beck, T. Posske, D. Crawford, E. Mascot, S. Rachel, R. Wiesendanger, and J. Wiebe,
Nat. Phys. \textbf{17,} 943 (2021).

\bibitem{VishwanathRMP}{N.-P. Armitage, E.-J. Mele, and A. Vishwanath, {Rev. Mod. Phys. \textbf{90,} 015001 (2018)}.}

\bibitem{XuReview}{Y. Xu, Front. Phys. \textbf{14,} 43402 (2019).}

\bibitem{Xu PRL2014}
Y. Xu, R.-L. Chu, and C.-W. Zhang,
Phys. Rev. Lett. \textbf{112}, 136402 (2014).

\bibitem{Xu PRL2015}
Y. Xu, F. Zhang, and C.-W. Zhang,
Phys. Rev. Lett. \textbf{115}, 265304 (2015).

\bibitem{HuPRL2014}
Y. Cao, S.-H. Zou, X.-J. Liu, S. Yi, G.-L. Long, and H. Hu,
Phys. Rev. Lett. \textbf{113}, 115302 (2014).

\bibitem{Xu2015PRLBKT}
Y. Xu and C. Zhang,
Phys. Rev. Lett. \textbf{114}, 110401 (2015).

\bibitem{Kamenev PRL2018}
X. Ying and A. Kamenev,
Phys. Rev. Lett. \textbf{121}, 086810 (2018).

\bibitem{Utsumi2004PRB}
H. Imamura, P. Bruno, and Y. Utsumi,
Phys. Rev. B \textbf{69}, 121303(R) (2004).

\bibitem{IBM1990NATURE}
D. M. Eigler and E. K. Schweizer, Nature \textbf{344}, 524 (1990).

\bibitem{heimes2015interplay}
A. Heimes, D. Mendler, and P. Kotetes,
New J. Phys. \textbf{17}, 023051 (2015).

\bibitem{Greiner PRL1997}
A. Greiner and L. Reggiani, T. Kuhn, and L. Varani
Phys. Rev. Lett. \textbf{78}, 1114 (1997).

\bibitem{Lutchyn PRB2016}
J. H. Zhang, Y. Kim, E. Rossi, and R. M. Lutchyn,
Phys. Rev. B \textbf{93}, 024507 (2016).
\end{thebibliography}
\end{document}